\documentclass[twocolumn,groupedaddress,showpacs,amsfonts,amssymb,amsmath]{revtex4}
\usepackage{graphicx}

\begin{document}

\bibliographystyle{plain}

\title{Nearest neighbor embedding with different time delays}

\author{Sara P. Garcia}
\email[]{spinto@itqb.unl.pt} \affiliation{Biomathematics Group,
Instituto de Tecnologia Qu\'imica e Biol\'ogica, Universidade Nova
de Lisboa, Rua da Quinta Grande 6, 2780-156 Oeiras, Portugal}

\author{Jonas S. Almeida}
\affiliation{Department of Biostatistics, Bioinformatics and
Epidemiology, Medical University of South Carolina, 135 Cannon
Street, Charleston, SC 29425, USA}

\date{\today}

\begin{abstract}
A nearest neighbor based selection of time delays for phase space
reconstruction is proposed and compared to the standard use of
time delayed mutual information. The possibility of using
different time delays for consecutive dimensions is considered. A
case study of numerically generated solutions of the Lorenz system
is used for illustration. The effect of contamination with various
levels of additive Gaussian white noise is discussed.\\
\begin{center}
(\copyright 2005 The American Physical Society,
\href{http://link.aps.org/abstract/PRE/v71/e037204}{http://link.aps.org/abstract/PRE/v71/e037204})
\end{center}
\end{abstract}


\maketitle

Reconstructing the phase space of a dynamical system from a time
series is a well-known mathematical result central to almost all
nonlinear time series analysis methods (see \cite{1} for a general
introduction). It is of paramount importance as it ensures that,
under certain generic conditions, such a reconstruction is
equivalent to the original phase space. This equivalence ensures
that differential information is preserved and allows for both
qualitative and quantitative analysis. Consider a smooth
deterministic dynamical system $s(t)=f(s(t_0))$, either in
continuous or discrete time, whose trajectories are asymptotic to
a compact $d$-dimensional manifold $\mathcal{A}$. When performing
$k$-dimensional measurements, where $k=1, \ldots, d$, a function
$\mathbf{x}_{(i)}=h[s(t=i \times \delta)]$ relates the states of
the dynamical system throughout time and a time series of measured
points, where $\mathbf{x}_{(i)} \in \mathbb{R}^{k}$, $i=1, \ldots,
n$; $n$ is the total number of sampled points, and $\delta$ is the
sampling time. As a consequence of our ignorance on the system, or
of limitations of the measurement apparatus, or simply because it
is too costly, $d$-dimensional measurements are typically not made
\cite{2,3}. In this report we will only address scalar
measurements, that is, $k=1$. Phase space reconstruction by time
delay embedding is a method of generating an $m$-dimensional
manifold that is equivalent to the original $d$-dimensional
manifold, by means of a matrix of delay-coordinate vectors.
Consider a column vector time series $\mathbf{x}_{(i)}$. Define an
$m$-dimensional matrix of delay-coordinate column vectors by
adding together displaced copies of the time series,
$\mathbf{X}=[\mathbf{x}_{(i)}, \mathbf{x}_{(i+\tau)}, \ldots,
\mathbf{x}_{(i+(m-1) \tau)}]$. Such matrix
$\mathbf{X}_{[n-(m-1)\times \tau,m]}$ is called an embedding
matrix, and two parameters need to be optimally estimated. The
first is the time delay $\tau$, which quantifies the time
displacement between successive delay-coordinate vectors. The
second is the embedding dimension $m$, which quantifies the number
of such delay-coordinate vectors. In this report we only address
the estimation of $\tau$, by suggesting a nearest neighbor based
procedure and comparing it to the standard use of time delayed
mutual information. Though in the limit of infinite data and
infinite precision $\tau$ may be set to any arbitrary value, a
balance between relevance and redundancy \cite{3} must be
accomplished for real data. When $\tau$ is too small, the elements
of the delay-coordinate vectors will mostly be around the
bisectrix of the phase space and, consequently, the reconstruction
will not be satisfactory. On the contrary, if $\tau$ is too large
the delay-coordinate vectors will become increasingly
uncorrelated, with the consequent loss of ability to recover the
underlying attractor. In addition, using a time delay larger than
necessary will render fewer data points for the reconstruction.
This may be particularly limiting for the study of biological
systems, where data sets are often not long. The most common
procedure for selecting $\tau$ is using the first minimum of time
delayed mutual information, as proposed by Fraser and Swinney
\cite{4}:
$I(x_{(i)},x_{(i+\tau)},\tau)=H(x_{(i)})+H(x_{(i+\tau)})-H(x_{(i)},x_{(i+\tau)})=\sum
p(x_{(i)},x_{(i+\tau)})\log_{2}\frac{p(x_{(i)},x_{(i+\tau)})}{p(x_{(i)})p(x_{(i+\tau)})}$,
where $H(x)$ is the Shannon entropy \cite{5}. Nonetheless its
widespread use, some drawbacks can be pointed out to this
selection criterion. The first is that probabilities are estimated
by creating a histogram for the probability distribution of the
data, which depends on a heuristic choice of number of bins, for
example, $\log_2$ of the total number of points \cite{6}.
Therefore, $I$ depends on the partitioning. The second drawback is
that it contains no dynamical information, which might be
incorporated by considering transition rather than static
probabilities, but such correction is usually not made \cite{7}.
The third is that the selection criterion presented by Fraser and
Swinney \cite{4}, though generalized to higher dimensions, was
established for two-dimensional embeddings \cite{3}, and does not
necessarily hold for higher dimensional embeddings, as shown
below. Finally, a fourth drawback \cite{3} is associated with the
fact that, when the purpose is solely to maximize statistical
independence \cite{4}, there is no obvious reason to choose the
first minimum over others. We propose an alternative measure for
selecting time delays, based on nearest neighbor estimations. This
nearest neighbor measure is inspired by the false nearest
neighbors algorithm proposed by Kennel \emph{et al.} \cite{8}.
With minimal assumptions, this measure is based solely on
topological and dynamical arguments documented by the data. We do
not address the estimation of $m$. The embedding theorem proposed
by Takens \cite{9} guarantees a solution. It states that if a map
from the original $d$-dimensional phase space $\mathcal{A}$, to
the reconstructed $m$-dimensional phase space is generic, when
$m\geq (2d+1)$ that map is a diffeomorphism on $\mathcal{A}$, that
is, an embedding, or a smooth one-to-one map with a smooth
inverse. This one-to-one property implies that if the system is
deterministic, distinct points on the attractor $\mathcal{A}$ are
mapped to distinct points under the embedding map \cite{2}.
Nevertheless, Takens result is only a sufficient condition
according to Kennel \emph{et al.} \cite{8}, who propose the use of
false nearest neighbors (F) as a criteria. Their algorithm
considers the ratio of Euclidean distances between a point and its
nearest neighbor, first on a $m$-dimensional and then on a
$(m+1)$-dimensional space. If the ratio is greater than a given
threshold, these points are referred to as F, that is, points that
appear to be nearest neighbors not because of the dynamics, but
because the attractor is being viewed in an embedding space too
small to unfold it. The procedure is repeated for all points in
the time series. As the fraction of F as a function of the
embedding space dimension decreases for deterministic systems,
when its value is zero, the underlying attractor is unfolded and
$m$ can be optimally estimated.

Considering the problem of optimally selecting time delays, we
will compare two different approaches. The first is a standard
procedure and uses the first minimum of the time delayed mutual
information to set $\tau$ for all dimensions \cite{4}. The second,
the one we propose, uses the first minimum of a nearest neighbor
measure to set the time delay for each dimension. Therefore, this
second procedure is iterative and introduces two novelties: using
a nearest neighbor based measure instead of the time delayed
mutual information, and using different time delays for
consecutive dimensions, as the standard use of the same $\tau$
value is an assumption out of convenience and not imposed by any
theoretical argument \cite{3}. In both cases, the embedding
dimension is estimated as the fraction of F decreases to zero. The
implementation of the standard procedure is described below. (i)
Consider an initial column vector time series $\mathbf{x}_{(i)}$.
For each $\tau$ being tested, $\tau = 1, \ldots, \frac{1}{10} n$,
build a temporary embedding matrix $\mathbf{T}=[\mathbf{x}_{(i)},
\mathbf{x}_{(i+\tau)}]$ out of two column vectors
$\mathbf{x}_{(i)}$ and $\mathbf{x}_{(i+\tau)}$. The upper limit
for $\tau$ is set arbitrarily. (ii) Estimate the time delayed
mutual information $I(\mathbf{x}_{(i)},\mathbf{x}_{(i+\tau)})$.
(iii) Select the first minimum from the profile of $I$ vs $\tau$,
which will be the optimal time delay for all dimensions (columns)
of the final embedding matrix $\mathbf{X}$. (iv) Estimate the
percentage of F (algorithm in \cite{8}) as a function of the
dimensionality of the embedding matrix. The optimal embedding
dimension is set when the fraction of F drops to $0$. As $\tau$ is
the same for all dimensions it will be referred to as a
\emph{fixed time delay}, and the final embedding matrix will be
$\mathbf{X}=[\mathbf{x}_{(i)}, \mathbf{x}_{(i+\tau)}, \dots,
\mathbf{x}_{(i+ (m-1)\tau)}]$. The implementation of the proposed
algorithm is as follows. (i) Consider an initial column vector
time series $\mathbf{x}_{(i)}$. For each $\tau$ being tested,
$\tau = 1, \ldots, \frac{1}{10} n$, build a temporary embedding
matrix $\mathbf{T}=[\mathbf{x}_{(i)}, \mathbf{x}_{(i+\tau)}]$.
(ii) For each two-dimensional point, that is, for each row in
matrix $\mathbf{T}$, estimate its (two-dimensional) nearest
neighbor. Calculate the Euclidean distance between them, $d_{E1}$.
(iii) Consider both points one sampling unit ahead and calculate
the new Euclidean distance between them, $d_{E2}$. (iv) Estimate
$d_{E2}/d_{E1}$ and save the number of distance ratios larger than
$10$. That fraction will be referred to as $N$. The threshold
value, though heuristically set, is justified by numerical studies
\cite{8} and has low parametric sensitivity. (v) Select the first
minimum from a profile of $N$ vs $\tau$, which will be the optimal
time delay for this first embedding cycle, $\tau_1$. We define an
\emph{embedding cycle} as each iteration [steps (i) to (vi)] that
adds another dimension to the embedding matrix. (vi) Estimate the
percentage of F. Save that value as a function of the
dimensionality of the temporary embedding matrix $\mathbf{T}$.
(vii) Consider now matrix $\mathbf{X}=[\mathbf{x}_{(i)},
\mathbf{x}_{(i+\tau_1)}]$ as the starting point for the second
embedding cycle. For each $\tau$ being tested, build a temporary
embedding matrix $\mathbf{T}=[\mathbf{x}_{(i)},
\mathbf{x}_{(i+\tau_1)}, \mathbf{x}_{(i+\tau)}]$. (viii) Repeat
steps (ii) to (vii), considering that points are now three- and
more dimensional, until the fraction of F drops to $0$. As there
will be a vector of $\tau$ values $[\tau_1, \dots, \tau_{(m-1)}]$,
this procedure is said to use \emph{different time delays}, and
the final embedding matrix will be $\mathbf{X}=[\mathbf{x}_{(i)},
\mathbf{x}_{(i+\tau_1)}, \mathbf{x}_{(i+\tau_2)}, \dots,
\mathbf{x}_{(i+ \tau_{(m-1)})}]$.

The Lorenz system \cite{10} of differential equations
$\dot{x}=\sigma(y-x)$, $\dot{y}=x(\rho-z)-y$, $ \dot{z}=xy-\beta
z$, with parameters $\sigma = 10, \rho = 28, \beta = 8/3$ will be
used as a case study. The equations were numerically integrated
with a 4-5th order Runge-Kutta algorithm and sampled at $\delta =
0.01$ intervals. Transients were removed. We will consider a first
data set, referred to as $L(X)$, which is the noise-free
$x$-coordinate of Lorenz system. We will also consider a second
data set, referred to as $L(X_\eta)$, consisting of the noise-free
$x$-coordinate of the Lorenz system contaminated with additive
Gaussian white noise of mean zero and variance 0.05, 1, 2, 3, or
5. For the major part of this report, noise of variance 1 will be
used. The other variances will be used later to further document
the effect of noise on the $\tau$ selecting procedures. Real
systems may also be contaminated with dynamical noise, though we
do not address such possibility here. Each data set includes a
total of 8000 points, $(\frac{1}{8})$th of which is plotted in
Fig. 1.

\begin{figure}[h]
\includegraphics{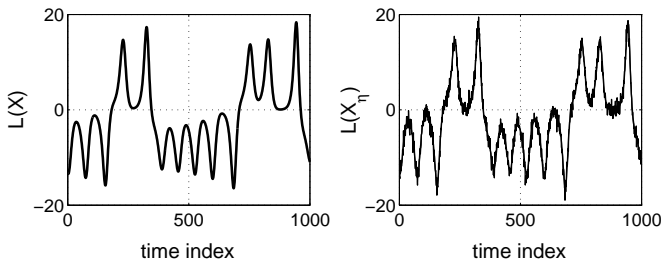}
\caption{\label{Fig. 1.}Data sets: noise-free $x$-coordinate of
the Lorenz system $[L(X)]$, and contaminated with additive
Gaussian white noise of mean 0 and variance 1 $[L(X_\eta)]$.}
\end{figure}

The profiles for selecting $\tau$ are displayed in Fig. 2 for the
first embedding cycle, and in Fig. 3 for the second embedding
cycle. Dashed lines represent $I$ profiles, while solid lines
represent $N$ profiles. Because the standard $\tau$ selecting
procedure uses the same time delay for all dimensions, only $N$
profiles are represented in Fig. 3. Displayed on the upper panels
are the profiles for $L(X)$ [Fig. 2(a) and Fig. 3(a)] and
$L(X_\eta)$ of variance 1 [Fig. 2(b) and Fig. 3(b)]. On the lower
panel, zoomed out versions of those same profiles are plotted to
document behavior beyond dynamic coupling. An arrow indicates the
global minimum of $N$ for the noise-free scenario [Fig. 2(a) and
(c), Fig. 3(a) and (c), solid line]. That same value can still be
identified for the first embedding cycle of the noisy data set
$L(X_\eta)$, though it is no longer a global minimum [Fig. 2(d),
solid line]. As explained previously, a value for $\tau$ is
selected from the first minimum of the $I$ profile for both $L(X)$
and $L(X_\eta)$ [Fig. 2(a) and (b), dashed line]; and a value for
$\tau_1$ is selected from the first minimum of the $N$ profile for
both $L(X)$ and $L(X_\eta)$ [Fig. 2(a) and (b), solid line]. For
the second embedding cycle, an asterisk indicates $\tau_1$, that
is, the value selected in the first embedding cycle, while a
circle indicates $\tau_2$, that is, the first minimum of this new
$N$ profile [Fig. 3(a)].

\begin{figure}[h]
\includegraphics{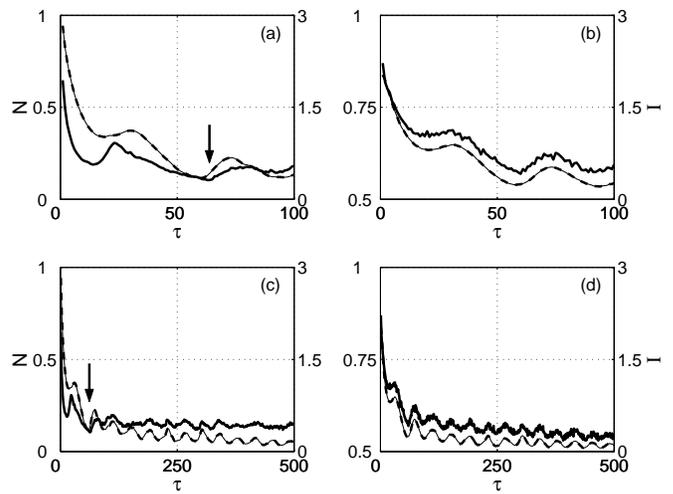}
\caption{\label{Fig. 2.}First embedding cycle profiles for $\tau$
selection from $I$ (dashed lines) and $N$ (solid lines). Upper
panel: (a) $L(X)$ and (b) $L(X_\eta)$ of variance 1. Lower panel:
zoomed out (c) $L(X)$ and (d) $L(X_\eta)$ of variance 1. An arrow
indicates the global minimum of $N$ for the noise-free scenario
[(a) and (c), solid line].}
\end{figure}

\begin{figure}[h]
\includegraphics{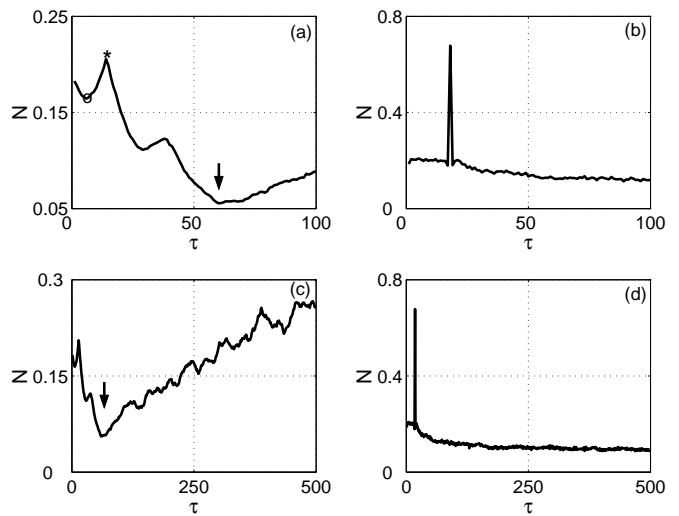}
\caption{\label{Fig. 3.}Second embedding cycle profiles for $\tau$
selection from $N$. Upper panel: (a) $L(X)$ and (b) $L(X_\eta)$ of
variance 1. Lower panel: zoomed out (c) $L(X)$ and (d) $L(X_\eta)$
of variance 1. An arrow indicates the global minimum of $N$ for
the noise-free scenario [(a) and (c)]. An asterisk indicates
$\tau_1$, while a circle indicates $\tau_2$ [(a)].}
\end{figure}

Three main conclusions can be drawn from examining the $\tau$
selecting profiles from both $I$ and $N$ for the first and second
embedding cycles. The first is that only $N$ retains the inverse
relationship with structure disclosure, that is, unlike $I$, $N$
values return to higher levels when the time delay is too long for
dynamical coupling to be retained [Fig. 2(c) and Fig. 3(c),
arrow]. This global minimum suggests an upper limit for the
efficient selection of $\tau$, beyond which statistical
independence reflects dynamic decoupling, and provides the
strongest argument for the use of $N$ over $I$. The effect of
noise will be discussed later. The second observation is that the
profiles for both embedding cycles are strikingly different,
indicating that reusing the time delay from a previous embedding
cycle is not an efficient procedure, as the $N$ profile peaks at
$\tau_1$ [Fig. 3(a), asterisk]. This peaking, an interesting but
presently unclear feature, was consistently observed for all
embedding cycles, and not only for the data sets analyzed here but
also for other systems, such as the R\"ossler attractor \cite{11},
not shown here for space constraints. The third conclusion refers
to the disruptive effect of additive noise, particularly clear in
Fig. 3(b). To further document such effect, profiles from $N$ for
the second embedding cycle and additive Gaussian white noise of
different variances are displayed in Fig. 4. The noise-free
scenario [Fig. 4(a), as in Fig. 3(a)] is compared to additive
Gaussian white noise of mean 0 and variance 0.05 [Fig. 4(b)], 1
[Fig. 4(c), as in Fig. 3(b)], 2 [Fig. 4(d)], 3 [Fig. 4(e)], and 5
[Fig. 4(f)]. All profiles peak exactly at $\tau_1$, as had been
previously observed in Fig. 3. Interestingly, the $\tau$ value
that is a global minimum in the noise-free scenario [Fig. 4(a),
arrow] is in most cases still identifiable. This feature may be a
helpful guideline, for the global minimum, though being a
suboptimal choice, sets the upper limit for the selection of
$\tau$ values.

\begin{figure}[h]
\includegraphics{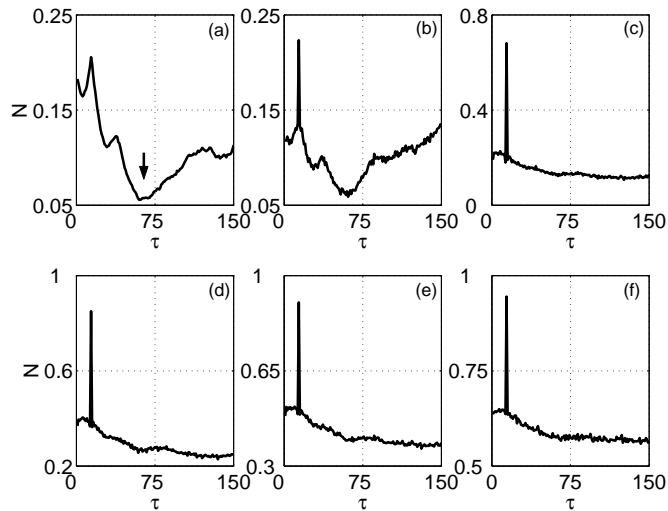}
\caption{\label{Fig. 4.} Second embedding cycle profiles for
$\tau$ selection from $N$: (a) $L(X)$ [as in Fig. 3(a)] and
$L(X_\eta)$ contaminated with additive Gaussian white noise of
mean 0 and variance (b) 0.05, (c) 1 [as in Fig. 3(b)], (d) 2, (e)
3, and (f) 5. An arrow indicates the global minimum of $N$ for the
noise-free scenario [as in Fig. 3(a)].}
\end{figure}

The second part of phase space reconstruction implies the
estimation of the embedding dimension. Fig. 5 documents the
profiles of F for increasing $m$ values, for the noise-free $L(X)$
data set. A dashed line represents the conventional use of $I$ and
fixed time delays, while a thick solid line represents using $N$
and selecting different time delays. The reconstruction using the
later is more efficient, in the sense that, though both $I$ and
$N$ suggest $m=3$ as the optimal embedding dimension, the
percentage of F when $m=2$ is lower for $N$. We have argued that
the global minimum of $N$ from the noise-free scenario would be an
upper limit for the efficient selection of $\tau$ values. A thin
solid line represents selecting the global minimum of $N$ as the
$\tau$ value for all embedding cycles, and it is clearly a
suboptimal choice. This further confirms the relevance of the
global minimum as a criterion for upper-limiting the region where
the selection of time delays should be made.

\begin{figure}[h]
\includegraphics{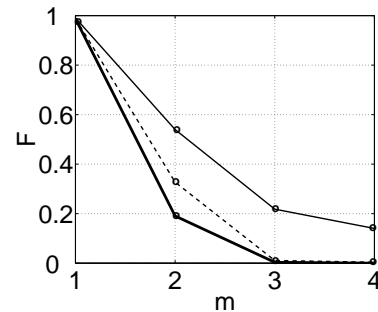}
\caption{\label{Fig. 5.}Profiles for $m$ selection from F for
$L(X)$: selecting a fixed $\tau$ from the first minimum of $I$
(dashed line), different $\tau$ values from the first minimum of
$N$ (thick solid line), and a fixed $\tau$ from the global minimum
of $N$ (thin solid line).}
\end{figure}

In summary, the nearest neighbor measure we propose, unlike mutual
information, retains the inverse relationship with structure
disclosure. This is an extremely useful feature for analyzing
noisy time series as it allows for the determination of an upper
limit to an efficient selection of time delays. Another extremely
important result is that the use of different time delays is more
efficient than the conventional use of a fixed time delay.

\begin{acknowledgments}  The authors thank S. Vinga and the referees
for insightful suggestions. This work was supported by grants
SFRH/BD/1165/2000 and POCTI/1999/BSE/34794 from Funda\c c\~ao para
a Ci\^encia e a Tecnologia, Portugal, and by the National Heart,
Lung and Blood Institute (NIH) Proteomics Initiative through
contract N01-HV-28181 (D Knapp, PI).
\end{acknowledgments}

\bibliography{GarciaAlmeidaNembedding}

\end{document}